# Voice Conversion with Diverse Intonation using Conditional Variational Auto-Encoder

*Soobin Suh[1], Dabi Ahn[2], Heewoong Park[1], Jonghun Park[1]*

[1]Seoul National University, Republic of Korea
[2]Kakao Brain, Republic of Korea

`soobin3230@snu.ac.kr, dabi.ahn@kakaobrain.com, hee188@snu.ac.kr, jonghun@snu.ac.kr`

## Abstract

Voice conversion is a task of synthesizing an utterance with target speaker's voice while maintaining linguistic information of the source utterance. While a speaker can produce varying utterances from a single script with different intonations, conventional voice conversion models were limited to producing only one result per source input. To overcome this limitation, we propose a novel approach for voice conversion with diverse intonations using conditional variational autoencoder (CVAE).

Experiments have shown that the speaker's style feature can be mapped into a latent space with Gaussian distribution. We have also been able to convert voices with more diverse intonation by making the posterior of the latent space more complex with inverse autoregressive flow (IAF). As a result, the converted voice not only has a diversity of intonations, but also has better sound quality than the model without CVAE.
**Index Terms**: voice conversion, variational autoencoder, intonations

## 1. Introduction

Imitating voice of other people, especially that of the celebrities, have been a topic of constant interest. Naturally, there has been many studies on building a voice conversion machine that converts the source speakers voice into that of a target speakers automatically. However, this model has a limit that it has a discriminative output which has only one intonation for one source speaker utterance due to the learned fixed variables. When reading a fairy tale aloud or giving a speech in front of a lot of audience, people utter exactly the same sentence with different intonations. This is not limited to the style among the different people but also to the individual's utterances. In this paper, we make the assumption that it needs stochastic information to generate a voice for a variety of styles that a person utters.

Recently, a variety of probabilistic deep generative models such as variational autoencoder (VAE) and generative adversarial networks (GAN) have been introduced and they have allowed to use stochastic information on various generation tasks. In [1] and [2], stochastic latent variables representing linguistic contents are learned from utterances by using VAE. In contrast, our CVAE based model learns latent space that representing intonations and uses linguistic content as condition. Since it is challenging to give a condition for a certain intonation and supervision is not easy to use, sampling a random value from the latent space is a promising approach for generating diverse intonations. As a result, in contrast to models that generate only a deterministic converted utterance, our model is able to generate utterances of a single speaker with diverse intonation.

Contributions of this paper are as follows:

- Within our knowledge, there has been no attempt to diversify intonations in voice conversion task. Using stochastic information, we have produced converted utterances with diverse intonations without performance degradation for unseen data.

- The proposed model showed better performance in terms of naturalness than the voice generated from the model learned without stochastic information.

## 2. Method

Our proposed method is based on a many-to-one model, which converts a utterance from any source speaker to a single target speaker's utterance. To extract speaker-independent linguistic features from source speaker's utterances and use those as conditions for generating target speaker's utterances, we structure the model into two modules, phoneme classifier and speech synthesizer. The speech synthesizer receives additional stochastic latent variable as input to generate diverse speech style. We train each modules separately in a way that phoneme classifier is trained before speech synthesizer.

### 2.1. Phoneme Classifier

Phoneme classifier converts each frame into a probability of phonemes. It is assumed that the logits before final softmax layer contain only linguistic features. Given pairs of a frame and corresponding one-hot encoded phoneme $\{x_{s,t}, y_{s,t}\}$ which subscript $s$ and $t$ represents a speaker and a timestep respectively, the model extracts linguistic feature as follows:

$$c_{s,t} = f(x_{s,t}, h_{t-1}) \quad (1)$$

where $f$ is the phoneme classifier, $h$ is a hidden state and $c_{s,t}$ is the probability of a phoneme at each time step $t$. In training, we use cross entropy loss to maximize likelihood as follows:

$$L_{PC} = \sum_{t=1}^{T} CELoss(c_{s,t}, y_{s,t}) \quad (2)$$

### 2.2. Speech Synthesizer

Speech synthesizer estimates target speaker's utterance spectrum at every timestep by inputting a probability of phonemes outputted from the phoneme classifier. To compare and show the effectiveness of our model, we propose a deterministic baseline model inspired by [3]. The baseline model is simply composed of two CBHG modules which is proposed by [4]. We then propose a CVAE-based model that uses linguistic features extracted from the phoneme classifier as a condition. CVAE is composed of an inference model which is an encoder, and a generative model which is a decoder. Given all frames of the target speaker $x_{tgt}$ and $c_{tgt}$ which is concatenation of phoneme probability features across all time steps, the inference model estimates the approximation posterior $z_{tgt}$ as follows:

$$\mu_{tgt}, \sigma_{tgt} = g(x_{tgt}, c_{tgt}) \quad (3)$$

$$z_{tgt} = \mu_{tgt} + \sigma_{tgt} * \epsilon \quad (4)$$

where $g$ is the inference model of CVAE, and $\epsilon$ is sampled from $N(0, I)$. Finally, using the fixed linguistic features $c_{tgt}$ and posterior $z_{tgt}$, the generative model $h$ generates spectrogram with diverse intonation.

$$\hat{x}_{tgt} = h(z_{tgt}, c_{tgt}) \quad (5)$$

In training phase, since optimizing CVAE directly from likelihood is intractable, we should maximize the lower bound of likelihood called ELBO [5]. According to the ELBO, the speech synthesizer's loss $L_{SS}$ containing reconstruction loss and KL divergence is calculated as follows:

$$L_{SS} = ||x_{tgt} - \hat{x}_{tgt}||_2^2 + \frac{1}{2}(\mu_{tgt}^2 + \sigma_{tgt}^2 - 2\log\sigma_{tgt} - 1) \quad (6)$$

### 2.3. Sample Generation

At inference time, the utterance of speaker that we want to convert is first passed onto phoneme classifier to get the linguistic features. Then, we sample a noise vector from the distribution $N(0, I)$. The noise vector and extracted linguistic feature from the classifier are passed through the speech synthesizer. To obtain converted utterances of diverse intonations, we repeatedly sample the random noise. We emphasize the predicted spectrogram by applying power of 1.2 to remove noisy sound, followed by Griffin-Lim Algorithm [6] to restore generated spectrogram to raw waveform.

### 2.4. Inverse Autoregressive Flow

To produce more diverse intonations, we should build a more flexible posterior than to sample $\epsilon$ from standard normal distribution. Inverse autoregressive flow (IAF) [7] is a kind of normalizing flow [8] that builds a flexible posterior distribution using a chain of iterative and invertible transformation from initial distribution. Using this method, the posterior probability density function can be easily computed by using the relation among $z$, $\mu$ and $\sigma$ at each time step. The final loss function using IAF is as below:

$$L_{SS} = ||x_{tgt} - \hat{x}_{tgt}||_2^2 + \frac{1}{2}(z_T^2 - \epsilon^2 - 2\sum_{t=1}^{T}\log\sigma_t) \quad (7)$$

## 3. Experiments & Results

### 3.1. Dataset

We trained the phoneme classifier on TIMIT corpus [9] which contains phonetically balanced 6,300 utterances and corresponding phones from 630 speakers of eight major dialects of American English. We trained the CVAE-based speech synthesizer on LJ Speech dataset which is one of Librispeech dataset [10].

### 3.2. Phoneme Classifier

We obtained top-1 evaluation accuracy of 73 percent in phoneme classification. Based on confusion matrix analysis, we discovered that the most frequently incorrect cases were 's'/'z', 'ah'/'aa', 'ae'/'eh', and 'ax'/'ix' which are the cases humans are confused too. We expect that data augmentation to be able to increase the accuracy even further, and we consider this to be a future work.

### 3.3. Diverse Intonation

We confirmed that the proposed model generated utterances with diverse intonations by plotting mel spectrograms and listening the samples. The results showed that if $\epsilon$ is sampled farther from $\mu$, the model generated an utterance with much more different intonation. However, if $\epsilon$ is sampled at the distribution over $3\sigma$, it resulted in loss of linguistic information, so that the synthesized utterance did not represent the correct sentence.

We also conducted an interpolation experiment on how intonation changes with $\epsilon$ changes. First, two utterances with the most different intonation were sampled to measure $\epsilon$ values, and then the experiment was performed while varying $\epsilon$ as follows.

$$\epsilon = \alpha\epsilon_1 + (1-\alpha)\epsilon_2 \quad (8)$$

where $\alpha$ is in the range of [0,1]. As a result, it was slightly and continuously changed according to $\epsilon$ when we plotted the mel spectrogram. However, when we heard the samples directly, it was hard to recognize the obvious variation with $\epsilon$ changes. Instead, we perceived it as a discrete change near a certain boundary. The changes of the mel spectrogram as gif files and the samples can be found at "https://soobinseo.github.io"

### 3.4. MOS quality

When comparing with the baseline model, the results of the proposed models had not only a variety of intonations but also much better quality. The 5-scale mean opinion score (MOS) was used to evaluate the naturalness of the generated utterance. We experimented on 3 models, which are baseline, vanilla VAE, and VAE with IAF, using two samples each from a male and a female. A total of 12 utterances generated from these three models and 8 utterances of ground truth were presented and evaluated in random order. As shown in Table 1, both models using VAE showed higher scores than the baseline model.

Table 1: *Mean Opinion Score (MOS)*

| Model | MOS score |
|---|---|
| Ground-truth (LJ Speech) | $4.83 \pm 0.07$ |
| Ground-truth (Arctic) | $4.24 \pm 0.18$ |
| Vanilla VAE | $2.70 \pm 0.25$ |
| VAE with IAF | $2.47 \pm 0.29$ |
| Non-VAE (baseline) | $2.23 \pm 0.27$ |

## 4. Conclusion

We presented a new voice conversion model that generates diverse intonations by using conditional variational auto-encoder. Unlike existing papers on voice conversion using variational auto-encoder that map the linguistic features to latent space, we give a diversity with intonation features as latent by pre-training the linguistic features and fixing it. Therefore, we could obtain not only voice with diverse intonations, but also better MOS performance.